\documentclass[showpacs,preprintnumbers,amsmath,amssymb]{revtex4}

\usepackage{epsfig}
\usepackage{graphicx}
\usepackage{dcolumn}
\usepackage{bm}
\newcommand{\be}{\begin{equation}}

\newcommand{\ee}{\end{equation}}
\newcommand{\bea}{\begin{eqnarray}}
\newcommand{\eea}{\end{eqnarray}}

\begin{document}

\preprint{BARI-TH 560/07}

\title{Phase diagram of a generalized  Winfree model}
\date{\today}
\author{F.~Giannuzzi}\email{floriana.giannuzzi@ba.infn.it}
\author{D.~Marinazzo}\email{daniele.marinazzo@ba.infn.it}
\author{G.~Nardulli} \email{giuseppe.nardulli@ba.infn.it}
\author{M.~Pellicoro}\email{mario.pellicoro@ba.infn.it}
\author{S.~Stramaglia}\email{sebastiano.stramaglia@ba.infn.it}
\affiliation{Dipartimento  Interateneo di Fisica, Universit\`a di
Bari, I-70126 Bari, Italy,} \affiliation{TIRES-Center of Innovative
Technologies for Signal Detection and Processing, Universit\`a di
Bari,  Italy,} \affiliation{I.N.F.N., Sezione di Bari, I-70126 Bari,
Italy}

\date{\today}

\begin{abstract}
We study the phase diagram of a generalized Winfree model. The modification is such that
the coupling depends on the fraction of synchronized oscillators, a situation which has
been noted in some experiments on coupled Josephson junctions and mechanical systems. We
let the global coupling $k$ be a function of the Kuramoto order parameter $r$ through an
exponent $z$ such that $z=1$ corresponds to the standard Winfree model, $z<1$ strengthens
the coupling at low $r$ (low amount of synchronization) and,  at $z>1$, the coupling is
weakened for low $r$. Using both analytical and numerical approaches, we find that $z$
controls the size of the incoherent phase region, and one may make the incoherent
behavior less typical by choosing $z<1$. We also find that the original Winfree model is
a rather special case, indeed the partial locked behavior disappears for $z>1$. At fixed
$k$ and varying $\gamma$, the stability boundary of the locked phase corresponds to a
 transition  that is continuous for $z< 1$ and first-order for $z>1$. This change in the nature of
the transition is in accordance with a previous study on a similarly modified Kuramoto
model.

\end{abstract}

\maketitle
\section{Introduction\label{intro}}Forty years ago in a
pioneering study  A. T. Winfree \cite{winfree67} (see also
\cite{winfree80}) introduced a mean field model to describe the
limit-cycle behavior of large populations of biological oscillators.
He discovered that systems of oscillators with randomly distributed
frequencies remain incoherent when the variance of the frequencies
is reduced, until a certain threshold is reached. Subsequently the
oscillators begin to synchronize spontaneously and become locked.

In its simplest form the model is defined by the set of equations
($i=1,...N$, $N\gg 1$ )
\begin{equation}
 \dot\theta_i(t)=\omega_i+\frac{k}N\sum_{j=1}^N
P(\theta_j) R(\theta_i)~. \label{eq1}
\end{equation} $\theta_i(t)$ is the
phase of the $i-$th oscillator; $\{\omega_i\}$ describes a set of
natural frequencies taken randomly from a distribution $g(\omega)$.
We shall assume below : $g(\omega)=1/2\gamma$ for $\gamma\in
[1-\gamma,1+\gamma]$, $g(\omega)=0$ otherwise; $ R(\theta_i)$ is the
sensitivity function giving the response of the $i-$th oscillator;
$P(\theta_j)$ is the influence function of the $j-$th oscillator. A
common choice is
\begin{equation}
 R(\theta)=-\sin\theta~,\hskip 1cm P(\theta)=1+\cos\theta~. \label{eqR}
\end{equation}
Despite its historical merits the Winfree's model has its own limitations. On one side it
is complex enough not to admit a full analytical treatment. On the other side it is not
sufficiently sophisticated as to allow the treatment of realistic systems. Limitations of
the former type were overcome by the work of Kuramoto \cite{kuramoto} (for a review
see\cite{stroreview}), who presented a model of oscillators  related to the Winfree's
model (it is the weak coupling limit of it) and analytically solvable in the mean field
approximation. The Kuramoto's approach generated an intense theoretical work
\cite{sakaguchi}, also motivated by the fact that the phenomenon of mutual
synchronization of coupled nonlinear oscillators is ubiquitous in nature, with
applications to neural networks, networks of cardiac pacemaker cells, populations of
fireflies and crickets \cite{winfree80}, \cite{stewart} as well as arrays of Josephson
junctions \cite{wiesenfeld}.

It must be also mentioned that, in spite of the complexity of the Winfree model, its
phase diagram  was object of investigation \cite{strogatz00,thesis} and in the
$(k,\,\gamma)$ plane  a rich structure was found.
 As to the
versatility of the Winfree's  model it can be mentioned that it can describe different
sets of pulse-coupled biological oscillators, see e.g.
\cite{walker,buck,peskin,angelini}. In view of its relevance it might be interesting to
look for extensions of the model that allow a different treatment of the couplings.
Hopefully these extensions should enlarge the class of physical instances where the model
can be usefully applied, for example experimental setups like arrays of Josephson's
junctions \cite{barbara,grib} or the crowd synchronization phenomenon on the Millenium
Bridge \cite{bridge}.

The purpose of this work is to generalize the Winfree model to the case of a global
coupling depending on the fraction of synchronized oscillators.  In a recent work
\cite{filatrella} a modification of the original Kuramoto model was presented. The
authors of \cite{filatrella} noted that the natural control parameter of the Kuramoto
model is a coupling strength, analogous to the $k$ parameter in \eqref{eq1}, independent
of the number of oscillators that are locked in frequencies. They suggested to generalize
the theoretical model by allowing a dependence of the coupling  on the number of locked
oscillators. The technical way to achieve this is to introduce a functional dependence on
the Kuramoto order parameter $r$, which is defined by the equation \be r(t)
e^{i\psi(t)}=\frac 1 N\sum_{j=1}^N e^{i\theta_j(t)}\ .\label{r} \ee Clearly in a locked
or partially locked phase $r$ does not vanish and its variability produces the desired
functional dependence. As the Kuramoto model is the averaged system of the Winfree model,
it is desirable to study the effect in this latter case. Therefore in this paper we study
the effect of an $r$-dependent coupling in the Winfree model, along the lines of
\cite{strogatz00,thesis}. Our main result is that in some cases the modification produces
an enlargement of the region, in the parameter space, where locking or partial locking of
the oscillators is possible.  It is worth noting that another interesting extension of
the Kuramoto model, where additional powers of the order parameter are introduced to
account for the dependence of the form of the coupling on its magnitude, has been
recently studied \cite{RP}. The plan of the paper is as follows. In the next section we
introduce the generalized Winfree's model, while in section 3 we describe how we obtain
the transition lines in the phase diagrams. Some conclusions are drawn in section 4.
\section{The generalized Winfree's model\label{model}}
 We will consider the
model defined by the set of equations
\begin{equation}
 \dot\theta_i(t)=\omega_i+\frac{k~r^{z-1}}N\sum_{j=1}^N
P(\theta_j) R(\theta_i) \label{eq2.1}
\end{equation} analogously to \eqref{eq1}, with $z$  a real
parameter.  It describes a set of  coupled non linear oscillators, with coupling constant
$k~r^{z-1}$. For $z=1$ the model reduces to the Winfree model of \eqref{eq1}.

For $N\to\infty$, the sum over all oscillators in~(\ref{eq1}) can be replaced by an
integral, yielding the following equation for the velocity  $v= \dot\theta$:
\begin{equation}
v(\omega,\theta,t)=\omega-\sigma(t)\sin\theta \label{eq3}
\end{equation} where
\begin{equation}
\sigma(t)= k\,r^{z-1}\int_0^{2\pi}\int_{1-\gamma}^{1+\gamma}\left(1+\cos\theta\right)
p\left(\theta,t,\omega\right) g(\omega) d\omega d\theta \;. \label{eq4}
\end{equation}Here $p\left(\omega,\theta,t\right)$ denotes the density of
oscillators with phase $\theta$ at time $t$. It satisfies the continuity equation
\be\frac{\partial p}{\partial t} =\,-\,\frac{\partial \left(p\,v\right)}{\partial
\theta}\label{continuity}\ee and the normalization condition\be\int_0^{2\pi} d\theta\,
p\left(\theta,t,\omega\right)~=~1~.\label{norma}\ee for all $\omega$ and any time.

The phase diagram of the model with $z=1$ was studied by Ariaratnam and Strogatz
\cite{strogatz00}. They found a rich structure, comprising: i) {\it locked phase} (Lk),
characterized by a common average frequency, ie a common value for the rotation number
$\rho_i=\lim_{t\to\infty}\theta_i(t)/t$; ii) {\it partial locking} (PL), characterized by
macroscopic fractions of locked and unlocked oscillators; iii) {\it incoherence} (In),
where no macroscopic fraction of oscillators is locked to a common frequency; iv) {\it
death} (Dt) characterized by $\rho_i=0$ for any $i$; v) {\it partial death} (PD), where
only a fraction the $\rho_i$ vanishes. Moreover they found several hybrid states that can
be seen as different realizations of the partial locking phase \cite{strogatz00}.

 We have numerically studied  the model \eqref{eq2.1} with various values of the parameter $z$
 in the interval $(0.5,\,2.0)$
  and various values of $N$, up to $N=1000$, starting from a random initial configuration.
  In general stability in the results is found after 500 time steps.
  Our results are qualitatively similar to those of the original Winfree's model
  ($z=1$), but the boundaries between the different phases depend on the actual
  value of $z$. Our numerical analysis is in general
  confirmed by the analytical study, see the next section.
  There is one case, the boundary $locking/partial$ $locking$,
   when  analytical results are not available, and the transition line must be evaluated only numerically.

The main outcome of the study is that the value of the parameter $z$
controls the size, in the phase diagram, of the incoherent phase:
one may make the incoherent behavior less typical by choosing $z<1$.
The original Winfree's model, corresponding to $z=1$, seems to be a
special case. As we describe in the following section, we find that
the partial locked region disappears for $z>1$ and that, at fixed
$k$ and varying $\gamma$,  the stability boundary of the locked
phase corresponds to a continuous transition for $z< 1$ and
first-order for $z>1$.

\section{Analytical and numerical results\label{analytic}}We shall define the transition
lines between different regions in the phase diagram starting from
areas where the solutions are stationary. Therefore we shall search
for solutions characterized by a density
$p_0\left(\omega,\theta\right)$ and a velocity
$v_0\left(\omega,\theta\right)$ independent
 of time. Clearly also $r$ and $\sigma$
 are time independent in \eqref{r} and \eqref{eq4}.

The continuity equation \eqref{continuity} has stationary solutions;
they satisfy $p_0v_0=C(\omega)$. From \eqref{eq3} we see that if
$\omega<\sigma$ one has the solution $C(\omega)\equiv 0$ and
therefore $v_0=0$. This implies that\be
p_0=\delta(\theta-\theta^*)\,,\hskip1cm
\sin\theta^*=\frac\omega\sigma\,, \label{death}\ee with the
condition\be\sigma\ge 1+\gamma\ .\label{death1}\ee The solution
\eqref{death}
 corresponds to the state of death
 (all the oscillators blocked at a fixed value of $\theta$). We shall assume $\theta^*\in(0,\pi/2)$,
 i.e. the result of the Winfree model \cite{strogatz00} as
  we have numerically tested that this result holds also for generic $z$.
  We shall discuss this solution in subsection
\ref{DPL}.

 If $\omega>\sigma$, then $C(\omega)\neq 0$ and  we get
\be
p_0\left(\omega\,,\theta\right)=\frac{C(\omega)}{\omega-\sigma\sin\theta}\
.\label{incoherence}\ee From the normalization condition one has \be
C(\omega)~=~\frac{\sqrt{\omega^2-\sigma^2}}{2\pi}\ .\ee In the
following we derive the stability boundaries between the phases of
the model, generalizing the methods in \cite{strogatz00,thesis}.

\subsection{Stability boundaries of the death phase\label{DPL}}

 From \eqref{r}  we get that, in general,
  \bea
r\sin\psi&=&\int_0^{2\pi} d\theta\int_{1-\gamma}^{1+\gamma} d\omega
g(\omega)\,\sin\theta\, p_0\left(\omega\,,\theta\right)\cr
r\cos\psi&=&\int_0^{2\pi}d\theta \int_{1-\gamma}^{1+\gamma} d\omega
g(\omega)\,\cos\theta\, p\left(\omega\,,\theta\right)\eea and in the
death state \bea r\sin\psi&=&\frac{1}{\sigma}\label{sin}\\
r\cos\psi&=&\int_{1-\gamma}^{1+\gamma}d\omega
\,g(\omega)\,\sqrt{1-\frac{\omega^2}{\sigma^2}}\label{cos}
 \eea that can be employed to determine $r$ and $\psi$.

 Let us use \eqref{eq4} to get
 \be \frac{\sigma}{k r^{z-1}}\,=\,1+r\cos\psi\ ,\label{ss}\ee
and adopt the definition \be G_\gamma(\sigma)=\sigma r^{1-z}\ . \ee
An explicit formula for the r.h.s of \eqref{ss} is obtained by  Eq.
\eqref{cos}. One gets \be 1+r\cos\psi\ = \ 1+\frac{\sigma}{4\gamma}
\left[\frac{1+\gamma}{\sigma}\sqrt{1-\left(\frac{1+\gamma}{\sigma}\right)^2}-
\frac{1-\gamma}{\sigma}\sqrt{1-\left(\frac{1-\gamma}{\sigma}\right)^2}+
\arcsin\frac{1+\gamma}{\sigma}-
\arcsin\frac{1-\gamma}{\sigma}\right]\,\equiv F_\gamma(\sigma) \
.\label{ss1}\ee The properties of  $F_\gamma(\sigma)$ were studied
in
 \cite{thesis}. For completeness we report here these results. It turns out that, as a function of $\sigma$
and for fixed  $\gamma$ , $F_\gamma(\sigma)$
   is a non-negative, increasing function, having the concavity
 down. From
\eqref{sin} we also have $r$ as a function of $\gamma$ and
$\sigma$:\be r=\sqrt{\frac
1{\sigma^2}\,+\,\left[F_\gamma(\sigma)-1\right]^2}\ . \ee
  We
distinguish the cases of large and small $\gamma$, the two ranges
being separated by a limiting value $\gamma_d\in(0,\,1)$ that
depends on $z$. For $z=1$, $\gamma_d=0.2956$ \cite{thesis}; let us
generalize this result using a procedure similar to  that of
\cite{thesis}. At the same time we will characterize the two regions
$\gamma\,<\,\gamma_d$ and $\gamma\,>\,\gamma_d$.

For  a fixed value of $\gamma$ such that $\gamma\,<\,\gamma_d$, the
two functions $G_\gamma(\sigma)/k$ and $F_\gamma(\sigma)$ can have
one, two or no intersection, depending on the value of $k$. The
value of $\sigma$ where the two curves are tangent, i.e.
$\sigma_d(\gamma)$, satisfies $\sigma_d(\gamma)>1+\gamma$ and is the
smallest value of $\sigma$ such that \eqref{ss} is satisfied.
Therefore it characterizes the boundary. We can use \eqref{ss} and
the tangency condition \be
\frac{G^\prime_\gamma(\sigma)}{k}=F^\prime_\gamma(\sigma)\
,\label{tangent}\ \ee  to extract  $\sigma_d(\gamma)$, getting rid
of $k$. In this way one gets the boundary in the form \be
k\,=\,\frac{G_\gamma(\sigma_d(\gamma))}{F_\gamma(\sigma_d(\gamma))}\
\hskip1cm{(\gamma<\gamma_d)}\  .\ee The procedure can be repeated
for various values of the parameter $z$ characterizing the
generalized Winfree's model.

\begin{figure}[ht]
\includegraphics[width=10cm]{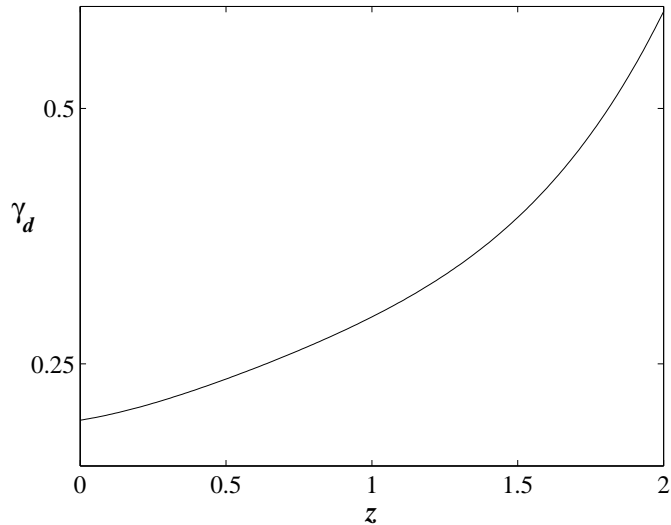}\caption{{\rm  The limiting value $\gamma_d$
as a function of $z$. The Winfree model is obtained for $z=1$.
 \label{fig1}}}\end{figure}

Increasing $\gamma$, $\sigma_d$ decreases and eventually it reaches the value $1+\gamma$.
For any  $\gamma$ such that $\gamma\,>\,\gamma_d$ the values $k$ are obtained by\be k\,
=\, \frac{G_\gamma(1+\gamma)}{F_\gamma(1+\gamma)}\ \hskip1cm{(\gamma>\gamma_d)}\  .\ee
\noindent The limiting value $\gamma_d$ is the solution of the equation \be
\frac{G_\gamma(1+\gamma)}{F_\gamma(1+\gamma)}\ =\
\frac{G^\prime_\gamma(1+\gamma)}{F^\prime_\gamma(1+\gamma)} \ .\ee The result, for
various values of $z$ in the interval $(0,2)$, is reported in Fig.\ref{fig1}.

The separation lines between the death region and the other phases (partial~death,
incoherence and locking-partial~locking) are reported in Fig. \ref{fig2} for four values
of $z$: 0.5, 1, 1.5 and 2. As stated above the analytical results are confirmed by the
numerical analysis.
 It can
be noted that the  boundaries of the death phase are almost independent on $z$. This
follows from the fact that the numerical values of $r$ are quite close to unity, as can
be seen expanding in the variable $ \gamma$: \be r(\gamma,\,\sigma)\,=\,1\,+\,
\frac{\gamma^2}{6(1-\sigma^2)}\,+\,
\frac{(31+9\sigma^2)\gamma^4}{360(1-\sigma^2)^3}+\,{\cal O}(\gamma^6)\ .\ee
\subsection{Transition Incoherence/Partial Death\label{LPL}}
Let us approach the boundary between these two phases from the the incoherence side. We
use \eqref{incoherence} in \eqref{eq4}. Since one must have $\sigma\le 1-\gamma$ the
boundary is obtained putting $\sigma=1-\gamma$. One has \bea r\cos\psi&=&0\label{cos2}\cr
r\sin\psi&=&\int_{1-\gamma}^{1+\gamma}d\omega
\,g(\omega)\,C(\omega)\int_0^{2\pi}d\theta\, \frac{\sin\theta}{\omega-\sigma\sin\theta}\
,\label{sin2} \eea so that \be r(\gamma,\,\sigma)\
=\,\frac{\sigma}{2\gamma}\left[f\left(\frac{1+\gamma}{\sigma}\right)-f\left(
\frac{1-\gamma}{\sigma}\right) \right] \label{r2},\ee where \be
f(x)=\frac{x^2}{2}-\frac{x\,\sqrt{x^2-1}}{2}\,+\,\frac 1 2\,\ln\left(x
+\sqrt{x^2-1}\right). \ee
\begin{figure}[ht]
\includegraphics[width=16cm]{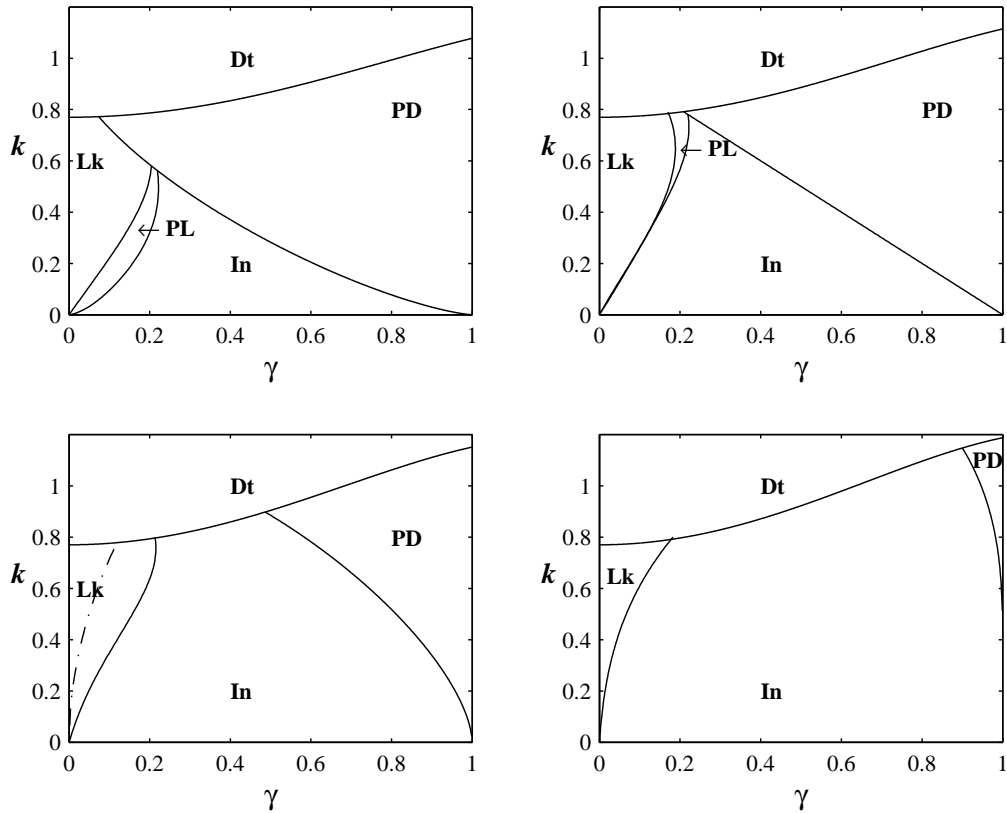}\caption{{\rm  Phase diagram of the generalized Winfree model. Results are for
$z=0.5$ (top-left), $z=1$ (top-right), $z=1.5$ (bottom-left) and
$z=2$ (bottom-right). Dt = death phase, PD = partial death, In =
incoherence, Lk = locked phase, PL = partial locking. The phase
diagram at $z=1$ coincides with that depicted in \cite{strogatz00}.
In the case $z=1.5$ the stability boundary between partial locking
and incoherence has been drawn dashed-dotted because it is not
observed in simulation (for $z>1$ the partial locking region is
absent, see the text).
 \label{fig2}}}\end{figure}

 It follows that
the boundary between the two regions is given by\be
k\,=\,(1\,-\,\gamma)\,\cdot
\Big[r(\gamma,\,1-\gamma)\Big]^{1-z}\,,\ee with $r$ given by
\eqref{r2}. Also these results are reported in Fig. \ref{fig2}. We
note that increasing $z$ the incoherent region increases while the
partial death region decreases.
\subsection{Transition Incoherence/Partial Locking\label{LPL}} In order
to determine the transition line we generalize the results of
\cite{thesis} to the case $z\neq 1$. One adds to the static solution
given in \eqref{incoherence}:
$p_0\left(\omega\,,\theta\right)=C(\omega)/(\omega-\sigma_0\sin\theta)$,
 a small time-dependent perturbation:\be
p\left(\omega\,,\theta\right)= p_0\left(\omega\,,\theta\right) \ + \
\epsilon\,\eta(\omega\,,\theta\,,t)\ .\label{incoherence2}\ee with
$\epsilon=0^+$
 and \be\int_0^{2\pi}d\theta\,\eta(\omega\,,\theta\,,t)\,=\,0\,.
 \ee
Similarly \be\sigma=\sigma_0\,+\,\epsilon\,\sigma_1 ,\ee with \be
\sigma_0=kr^{z-1}\label{sigma0}\ee and \be\sigma_1(t)=kr^{z-1}\int\,d\omega
\,g(\omega)\int_0^{2\pi}d\theta\,\eta(\omega\,,\theta\,,t), \ee so that \be
v=v_0-\epsilon\sigma_1(t)\sin\theta\  ,\ee with $v_0=\omega-\sigma_0\sin\theta$. From the
continuity equation one has, at first order in $\epsilon$: \be\frac{\partial
\eta}{\partial t}\,+\,\frac{\partial(\eta
v_0)}{\partial\theta}\,=\,\sigma_1(t)\,\frac{\omega\,C(\omega)\cos\theta}{v_0^2}.\ee
 Searching solutions in the form\be\eta=
 e^{\lambda t}h(\omega\,,\theta),\ee one finds\be
 \lambda h \,+\,\frac{\partial(h
v_0)}{\partial\theta}\,=\,A\,\frac{\omega\,C(\omega)\cos\theta}{v_0^2},\ee with \be
\label{A}A=kr^{z-1}\int\,d\omega \,g(\omega)\int_0^{2\pi}d\theta\,h(\omega\,,\theta), \ee
whose solution is \cite{thesis} \be
h(\omega\,,\theta)\,=\,\frac1{v^2_0}\Big(a\,+\,b\cos\theta\,+\,c\sin\theta\Big), \ee
with\be a=\,-\,A\,\frac{\sigma_0\omega
C(\omega)}{\lambda^2+\omega^2-\sigma_0^2}\,,~~~~~~~~~b=\,-\,\frac{\lambda
a}{\sigma_0}\,,~~~~~~~~~c=\,-\,\frac{\omega a}{\sigma_0}\ .\ee Therefore from \eqref{A}
one gets \bea
\sigma_0^2&=&\sigma_0\,I_{\sigma_0}\label{s1}\\
\sigma_0&=&k\,r^{z-1}\label{s2}\eea with \be I_{\sigma_0}=\int
d\omega \,g(\omega) \,\omega\,\lambda\,\frac{\omega^2\,-
\,\sqrt{\omega^2-\sigma_0^2}}{\lambda^2+\omega^2-\sigma^2_0}\label{Is}\ee
and, as in previous equation \eqref{r2} , \be
r=\,\frac{\sigma_0}{2\gamma}\left[f\left(\frac{1+\gamma}{\sigma_0}\right)-
f\left( \frac{1-\gamma}{\sigma_0}\right) \right]\ .\ee
\begin{figure}[ht]
\includegraphics[width=16cm]{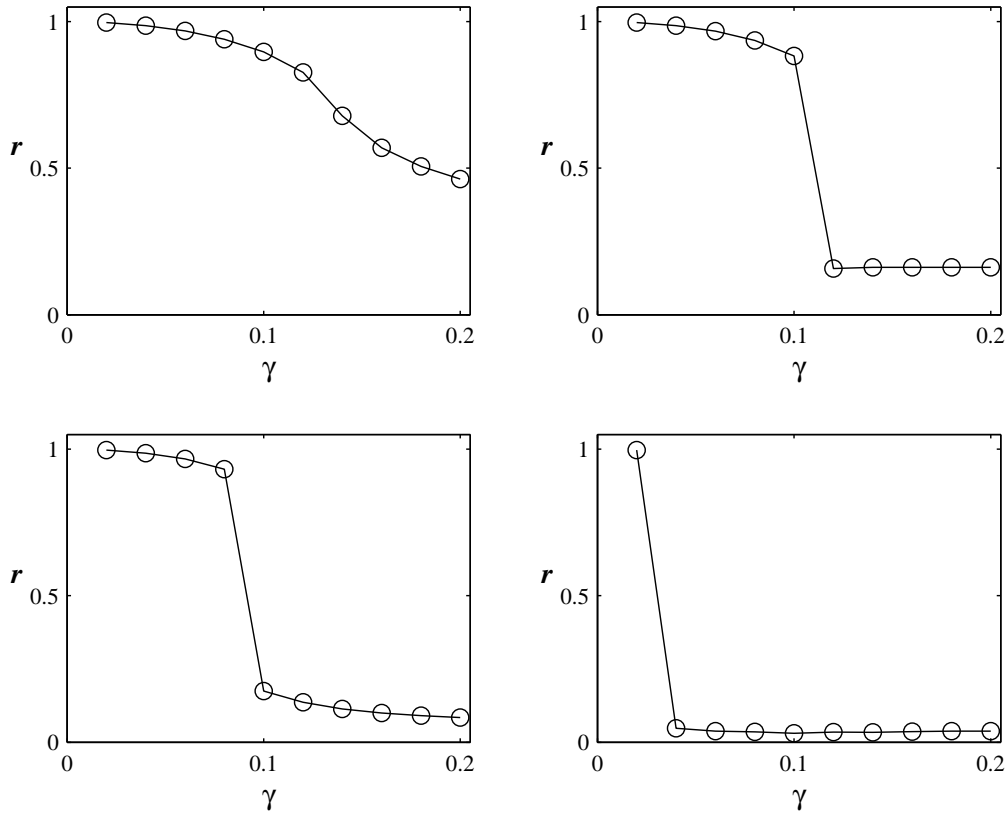}\caption{{\rm  The order parameter (time averaged) $r$ versus $\gamma$,
for low $\gamma$ and fixed $k=0.4$, is reported for $z=0.5$
(top-left, continuous transition between locked phase and partial
locked phase) , $z=1$ (top-right)  $z=1.5$ (bottom-left) and $z=2$
(bottom-right). In the last three cases there is a first order
transition between the locked phase and the incoherence phase.
 \label{fig3}}}\end{figure}

The transition line is obtained by taking the limit $\Re(\lambda)\to 0^+$. Approximate
results can be obtained performing a perturbative expansion for small $\gamma$, taking
into account that in this limit also $\sigma_0\to 0$.
 At the
lowest order in $\gamma$ we can write $\lambda=\,i\,+\,\lambda_1$
with $\lambda_1$ real. A better approximation can be obtained going
up to the fifth order. The result is identical
 to the one found in \cite{strogatz00}, \cite{thesis} for $z=1$, with the substitution
 $k\to\sigma_0$. At the lowest order in $\gamma$ one gets
 $\displaystyle r=\frac{\sigma_0}{2}$ and therefore from
\eqref{s2}\be \sigma_0=\,2\,\left(\frac k
2\right)^{\displaystyle\frac 1{2-z}}~,\ee which shows that there are
no solutions for $z>2$. From \eqref{Is} one
finds\be\sigma_0=\frac{8\gamma}{\pi}\left(1+\frac{16\gamma^2}{\pi^2}
\,+\,\frac{16(\pi^2+80)\gamma^4}{\pi^4}\right)\,+\,O(\gamma^7)\label{sigmazero}
\ee and $k$ is given by \be
k=2\left(\frac{\sigma_0}{2}\right)^{2-z}\ .\label{transition}\ee In
Fig. \ref{fig2} the separation line between $incoherence$ and
$partial$ $locking$ is computed using the exact expression of Eqns.
\eqref{s1}, \eqref{s2}, and \eqref{Is}. The approximate formula
based on the expansion \eqref{sigmazero} is valid within $4$ \% for
values of $\gamma$ not larger than $\simeq 0.21$ and $z=1$; for
$z=0.5$ the validity is within $4$ \% for $\gamma< 0.19$. It should
be noted that for $z>1$ moving from the right to the left at fixed
$k$ in the diagrams,  one encounters the $locking$ region
 before reaching the $partial$ $locking$ phase. Therefore, for $z>1$ the partial locking region is basically absent,
 which means that, starting from a random initial configuration, the system never reaches a partial locking
 state. We remark that the occurrence of this phenomenon does not depends on the choice
 of $g(\omega)$ uniform, indeed we verify that it holds also for a Lorentzian
 distribution. We refer the reader to \cite{pazo} for a discussion about the role of the
 shape of the distribution of frequencies in the Kuramoto model.
 \subsection{Transitions locking/partial~Locking ($z<1$) and locking/incoherence ($z>1$) \label{LPL}} To
derive the boundary from the locking to the partial~locking phase, one should define the
latter. This characterization can be only heuristic, given the composite nature of the
partial locked phase. We have obtained the transition line by the numerical solution of
(\ref{eq2.1}), using $N=800$ oscillators and $T=1000$ time units. To study the stability
of the locked phase one should distinguish two cases, $z< 1$ and $z> 1$. In the case
$z<1$ we find a continuous transition from the locked phase towards the partial locking
phase in which the time average of the order parameter decreases continuously between two
limit values (see figure \ref{fig3}). For $z> 1$ there is a first order transition from
the locked phase to the incoherent phase as the time average of the order parameter $r$
jumps from $r \sim 1$ (locked phase) to $r < 0.1$ (incoherent phase), as shown in figure
\ref{fig3}. In order to derive the curves limiting the locked phase, drawn in
fig.\ref{fig2}, we fix the coupling constant $k$;  for each value we consider the plot of
$r$ versus $\gamma$ and (i) for $z>1$ find the value of $\gamma$ at which the jump occurs
(ii) for $z< 1$ find the value of $\gamma$ at which the curve has a flex point. The
boundary line of the locked phase slightly depends on $z$ (in the locked phase $r\approx
1$): as $z$ decreases the locked phase region is slightly enlarged.

\section{Conclusions}
We have presented a modification of the Winfree model to account for
effective changes in the coupling constant among oscillators, as
suggested by experiments on Josephson junctions and mechanical
systems. The modification can be parametrized by a real number $z$.
 The case $z=1$ corresponds to the Winfree model; $z<1$ leads to a
coupling which decreases as the order parameter $r$ increases, thus
enforcing the coupling at low $r$ (low amount of synchronization);
at $z>1$ the coupling increases with $r$, i.e. the coupling is
weakened at low $r$. Using both an analytical approach and numerical
simulations we have outlined the phase diagram of the model as $z$
varies.

As figure \ref{fig2} clearly shows, the death phase region is almost
independent on $z$, whilst the region of incoherence is strongly
influenced by this parameter: for $z<1$ it shrinks, as the effective
coupling is  strengthened at a low amount of synchronization,
 whereas it widens at
$z>1$ at the expenses of the partial death region.

As far as the partial locking phase is concerned, we find that it
disappears at $z>1$ thus leading to the following phenomenon. At low
$k$, as $\gamma$ is increased the system leaves the locking phase
through a continuous transition (in the order parameter $r$) for
$z<1$, whilst for $z>1$ the system undergoes a discontinuous
transition while leaving the locked phase.  This happens because for
$z<1$ the partial locking phase separates the locking and the
incoherence phases, whereas for $z>1$ the transition is directly
onto the incoherence phase. The standard case $z=1$, hence, appears
to be rather special. It is worth noting that a similar change in
the nature of the transition was noticed in the generalized Kuramoto
model \cite{filatrella}, and a discontinuous transition was
experimentally seen in the synchronization of over-damped Josephson
junctions \cite{grib}, where physically the parameter $z$
corresponds to the degree of feedback provided by a coupling
resonator. Our results suggest strategies to control incoherent
behavior in systems of interacting oscillators with coupling
depending on the fraction of synchronized sub-units.

\vskip 2 cm \noindent The authors are indebted with S.H. Strogatz
and J.T. Ariaratnam for a valuable correspondence and for providing
copy of \cite{thesis}.


\begin{thebibliography}{99}
\bibitem{winfree67} A.~T. Winfree, J.~Theor. Biol. {\bf 16}, 15 (1967).
\bibitem{winfree80} A.~T. Winfree, {\em The Geometry of Biological
Time\/} (Springer, New York, 1980).
\bibitem{kuramoto}Y. Kuramoto, in
 {\it International Symposium on Mathematical
 Problems in Theoretical Physics}, Vol. 39 of {\it
 Lecture Notes in Physics},
  edited by H. Araki (Springer-verlag,
 Berlin, 1975); {\it Chemical Oscillations,
 Waves and Turbulence} (Springer-verlag,
 Berlin, 1984).

\bibitem{stroreview}S.~H. Strogatz, Physica~D {\bf 143}, 1 (2000).

\bibitem{sakaguchi}
H. Sakaguchi and Y.~Kuramoto, Prog. Theor.
Phys. {\bf 76}, 576 (1986);Y.~Kuramoto and I.~Nishikawa, J.~Stat.
Phys. {\bf 49}, 569 (1987); S.~H. Strogatz and R.~E. Mirollo,
J.~Stat. Phys. {\bf 63}, 613 (1991); L.~L. Bonilla, J.~C. Neu, and
R.~Spigler, J. Stat. Phys. {\bf 67}, 313 (1992); H.~Daido, Phys.
Rev. Lett. {\bf 73}, (1994) 760; J.~D. Crawford, J.~Stat. Phys. {\bf
74}, 1047 (1994).

\bibitem{stewart}S.H. Strogatz and I. Stewart, Sci.
Am. (Int. Ed.) {\bf 269} (6), 102 (1993).

\bibitem{wiesenfeld}K.Wiesenfeld, P. Colet and S.
H. Strogatz, Phys. Rev. Lett. {\bf 76}, 404 (1996); K.Wiesenfeld, P.
Colet and S. H. Strogatz, Phys. Rev. E {\bf 57}, 1563 (1998); M.
Dhamala and K. Wiesenfeld, Phys. Lett. A {\bf 292}, 269 (2002).

\bibitem{strogatz00}J.T. Ariaratnam and  S.H. Strogatz, Phys. Rev. Lett.
 {\bf 86}, 4278
 (2001).
\bibitem{thesis}J.T. Ariaratnam, PhD Thesis, unpublished.

\bibitem{walker} T.~J. Walker, Science {\bf 166}, 891
(1969); E.~Sismondo, {\em ibid.}, {\bf 249}, 55 (1990).
\bibitem{buck} J.~Buck, Quart. Rev. Biol. {\bf 63}, 265 (1988).
\bibitem{peskin} C.~S. Peskin, {\em Mathematical Aspects of Heart
Physiology\/} (Courant Inst. Math. Sci., New York, 1975); D.~C.
Michaels, E.~P. Matyas and J.~Jalife, Circ. Res. {\bf 61}, 704
(1987).
 \bibitem{angelini}L. Angelini et al., Phys. Rev. E {\bf 69}, 061923
 (2004).
\bibitem{barbara}P. Barbara, A. B. Cawthorne, S. V. Shitov and C. J. Lobb. Phys.
 Rev. Lett. {\bf 82}, 1963 (1999).
\bibitem{grib} A.N. Grib, P. Seidel, and J. Scherbel, Phys. Rev. {\bf B 65}, 94508
  (2002).  \bibitem{bridge}S. H. Strogatz, D. M.Abrams, A. McRobie, B.
  Eckhardt and E. Ott, Nature, {\bf 438}, 43 (2005).
\bibitem{filatrella}G. Filatrella, N. F. Pedersen and K Wiesenfeld, Phys. Rev. E {\bf 75}, 017201 (2007); Physica {\bf C 437}, 65 (2006).
\bibitem{RP}  M. Rosenblum and A. Pikovsky,
Phys. Rev. Lett. {\bf 98}, 064101 (2007).
\bibitem{pazo}  D. Paz\'o, Phys. Rev. E {\bf 72},
046211 (2005).
\end{thebibliography}
\end{document}